\documentclass{PoS}

\usepackage{graphicx}
\usepackage[nice]{nicefrac}
\usepackage{amsmath,bm}
\usepackage{subfig}

\newcommand\T{\rule{0pt}{2.6ex}}

\title{Form factors for $B$ and $B_s$ semileptonic decays with NRQCD/HISQ quarks}
\ShortTitle{$B$ and $B_s$ semileptonic decays with NRQCD/HISQ quarks}

\author{
Chris~M.~Bouchard$^{\, *\, a}$,
G.~Peter~Lepage$^{\, b}$,
Chris~J.~Monahan$^{\, c}$,
Heechang~Na$^{\, d}$,
Junko~Shigemitsu$^{\, a}$
\hphantom{\speaker{C.M.~Bouchard}}
\\ \\
\llap{$^a$}Department of Physics, The Ohio State University, Columbus, OH 43210, USA\\
\llap{$^b$}Laboratory of Elementary Particle Physics, Cornell University, Ithaca, NY 14853, USA\\
\llap{$^c$}Department of Physics, College of William and Mary, VA 23187-8795, USA\\
\llap{$^d$}Argonne Leadership Computing Facility, Argonne National Laboratory, Argonne, IL 60493, USA}
\author{HPQCD Collaboration\\
\email{bouchard.18@osu.edu}}

\abstract{We discuss preliminaries of a calculation of the form factors for the semileptonic decays $B\to\pi \ell\nu$, $B_s\to K\ell\nu$,  and $B\to K\ell^+\ell^-$. We simulate with NRQCD heavy and HISQ light valence quarks on the MILC $2+1$ dynamical asqtad configurations. The form factors are calculated over a range of momentum transfer to allow determination of their shape and the extraction of $|V_{ub}|$. Additionally, we are calculating ratios of these form factors to those for the unphysical decay $B_s\to\eta_s$. We are studying the possibility of combining these precisely determined ratios with future calculations of $B_s\to\eta_s$  using HISQ $b$-quarks to generate form factors with significantly reduced errors.}

\FullConference{The 30th International Symposium on Lattice Field Theory - Lattice 2012,\\
		June 24-29, 2012\\
		Cairns, Australia}

\begin{document}

\section{Motivation}
\label{sec-Motivation}
We are improving upon our previous $B\to\pi\ell\nu$ calculation~\cite{HPQCD:2006} in several ways, including the use of:  $b$-quark smearing; HISQ light valence-quarks with random wall sources~\cite{HPQCD:2010}; better scale-determination~\cite{HPQCD:scale}; fitting advances ({\it e.g.} simultaneous fits to multiple separation times); and the $z$-expansion~\cite{zexp}.  The calculation will also benefit from improved experimental data~\cite{BPilvexpt} which, when combined with lattice results, determines $|V_{ub}|$.

In parallel, we are studying the $B_s \to K \ell \nu$ decay.  In combination with planned measurements~\cite{expt:BsKlv}, this will provide an additional exclusive determination of $\big| V_{ub} \big|$.  Not yet studied on the lattice, this decay has a heavier spectator quark than $B\to \pi \ell \nu$ and should have reduced errors.

We are also studying $B\to K\ell^+\ell^-$, where the flavor-changing neutral current $b\to s$ provides a probe of new physics ({\it cf.} Ref.~\cite{BKll:NP}).  There are existing~\cite{expt:BKll} and promised~\cite{Bfacs:2010} experimental results for this decay, but few unquenched lattice calculations~\cite{lat:BKll}.

Additionally, we are investigating the possibility of using the unphysical $B_s \to \eta_s$ decay to build ratios of form factors using NRQCD $b$-quarks in which the leading sources of error largely cancel.  This ratio could be combined with a future calculation of $B_s\to \eta_s$ using a HISQ $b$-quark to yield form factors with greater precision, {\it ie.}
\begin{equation}
\left. \frac{f(B\to \pi \ell \nu)}{f(B_s \to \eta_s)} \right|_{{\rm NRQCD}\ b} \times \left. f(B_s\to\eta_s) \right|_{{\rm HISQ}\ b},
\end{equation}
analogous to the recent HPQCD work on $B$ and $B_s$ decay constants~\cite{HPQCD:2012}.

\section{Calculation}
The Standard Model $(V-A)^\mu$ weak interaction responsible for the $b\to u$ transition results in hadronic matrix elements $\langle X | V^\mu|B_q\rangle$, parameterized via form factors
\begin{equation}
\langle X | V^\mu | B_q \rangle = f_+^{B_qX}(q^2) \left( p_{B_q}^\mu +p_X^\mu - \frac{ M_{B_q}^2 - M_X^2 }{ q^2 }\,q^\mu \right) + f_0^{B_qX}(q^2)\frac{ M_{B_q}^2 - M_X^2 }{ q^2 }\, q^\mu,
\end{equation}
where $q^\mu \equiv p_{B_q}^\mu - p_X^\mu$.  We recast these form factors in terms of lattice-convenient form factors,
\begin{equation}
\langle X | V^\mu | B_q \rangle = \sqrt{2M_{B_q}} \left[ \frac{p_{B_q}^\mu}{M_{B_q}}\ f_\parallel^{B_qX}(q^2) + p_\perp^\mu\  f_\perp^{B_qX}(q^2) \right],
\end{equation}
where $p_\perp^\mu \equiv p_X^\mu - (p_X\cdot p_{B_q})\nicefrac{p_{B_q}^\mu}{M_{B_q}^2}$.  In the $B_q$-meson rest frame, the form factors are simply related to the temporal and spatial components of the hadronic vector matrix elements,
\begin{eqnarray}
\langle X | V^0 | B_q \rangle &=& \sqrt{2M_{B_q}}\ f_\parallel^{B_qX}(q^2) \nonumber \\
\langle X | V^k | B_q \rangle &=& \sqrt{2M_{B_q}}\ p_X^k\ f_\perp^{B_qX}(q^2).
\end{eqnarray}
We calculate the components of the hadronic vector matrix elements and, from them, construct the form factors $f_{+,\,0}^{B_qX}(q^2)$ for the decays listed in Sec.~\ref{sec-Motivation}.  Ultimately, the form factors are related to experimentally measured differential decay rates\footnote{Eq.~(\ref{eq-diffDR}) neglects final state lepton masses.}
\begin{equation}
\frac{d\Gamma^{B_qX}}{dq^2} = \frac{G_F^2 \big| V_{ub} \big|^2}{ 192 \pi^3 M_{B_q}^3} \left[  \left( M_{B_q}^2 + M_X^2 - q^2 \right)^2 - 4M_{B_q}^2M_X^2 \right]^{3/2} \big| f_+^{B_qX}(q^2) \big|^2,
\label{eq-diffDR}
\end{equation}
where experimental and lattice results are combined to determine $\big| V_{ub}\big|$.  The Standard Model suppressed $b\to s$ transition in $B\to K\ell^+\ell^-$ opens the door for potentially discernible new physics contributions.  The search for new physics in this decay requires the tensor form factor, related to the $(k0)$-component of the hadronic tensor matrix element
\begin{equation}
 \langle K | T^{k0} | B\rangle = \frac{ 2 M_B p^k_K }{ M_B+M_K }\  if_T^{BK}(q^2).
\end{equation}

\subsection{Generating Correlator Data}
\begin{figure}[t]
\vspace{0.0in}
\centering
\hspace{0.0in}  
{\scalebox{0.9}{\includegraphics[angle=0,width=0.7\textwidth]{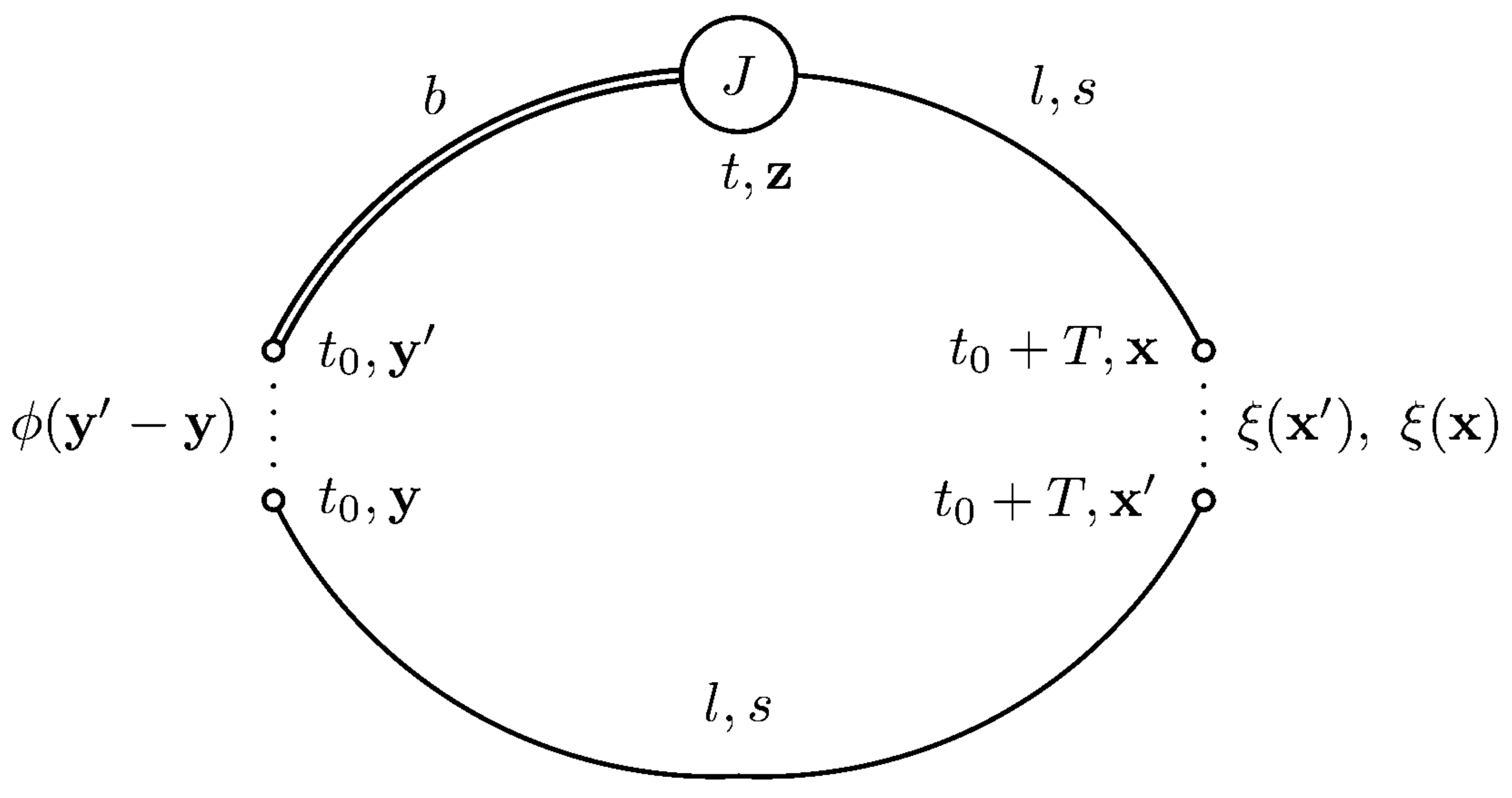}}}
\caption{Setup for three-point correlator data generation.}
\vspace{0.0in}
\label{fig-feyndiag}
\end{figure}

\begin{table}[t]
\vspace{3mm}
\centering
\begin{tabular}{ccccccc}
\hline\hline	
	\T ensemble   	& $\approx a$ [fm]	& $m_l({\rm sea})/m_s({\rm sea})$	& $N_{\rm conf}$	& $N_{\rm tsrc}$	& $L^3\times N_t$	& $T$	\\
	\hline
	\T C1 	     	& 0.12			& 0.005/0.05			& 1200		& 2			& $24^3 \times 64$	& 12 -- 15	\\
	\T C2		& 0.12			& 0.01/0.05			& 1200		& 2			& $20^3 \times 64$	& 12 -- 15 	\\
	\T C3		& 0.12			& 0.02/0.05			& 600		& 2			& $20^3 \times 64$	& 12 -- 15	\\ 
	\T F1 	     	& 0.09			& 0.0062/0.031			& 1200		& 4			& $28^3 \times 96$	& 21 -- 24	\\
	\T F2 	     	& 0.09			& 0.0124/0.031			& 600		& 4			& $28^3 \times 96$	& 21 -- 24	\\
\hline\hline
\end{tabular}\caption{Left to right:  ensemble, lattice spacing, light and strange sea-quark mass, number of configurations, number of source times, volume, and separation between parent and daughter mesons.}
\label{tab-ens}
\end{table}
Ensemble averages are performed using the MILC $2+1$ asqtad gauge configurations~\cite{MILC:2010} listed in Table~\ref{tab-ens}.  The valence quarks in our simulation are NRQCD~\cite{HPQCD:1992} $b$-quarks, tuned in Ref.~\cite{HPQCD:2012}, and HISQ~\cite{HPQCD:2007} light and strange quarks, whose propagators were generated in previous works~\cite{HPQCD:2010, HPQCD:2011}.
Working in the parent meson rest frame, a sequential propagator is built from NRQCD $b$ and spectator HISQ quarks.  The $b$-quark smearing function $\phi({\bf y}'-{\bf y})$ is either a delta function or Gaussian, specified by indices $\alpha,\beta$ in Eqs.~(\ref{eq-B2pt}, \ref{eq-3pt}), and is introduced by the replacement $\sum_{\bf y} \to \sum_{{\bf y},{\bf y}'}\phi({\bf y}'-{\bf y})$.  The spectator source includes a U(1) phase $\xi({\bf x}')$.  The daughter quark, with U(1) phase and momentum insertion at ${\bf x}$, is tied to the sequential quark propagator, with $\sum_{{\bf x}}$ in Eqs.~(\ref{eq-B2pt} - \ref{eq-3pt}) accomplished via random wall sources, {\it ie.} $\sum_{{\bf x}} \to \sum_{{\bf x},{\bf x}'}\xi({\bf x})\xi({\bf x}')$.
\begin{eqnarray}
C^{\alpha\beta}_{B_q}(t_0,t) &=& \frac{1}{L^3} \sum_{{\bf x}, {\bf y}}  \langle \Phi^\beta_{B_q}(t,{\bf y})\ \Phi^{\alpha\dagger}_{B_q}(t_0,{\bf x}) \rangle \label{eq-B2pt} \\
C_X(t_0,t;{\bf p}) &=& \frac{1}{L^3} \sum_{{\bf x}, {\bf y}}  e^{i\,{\bf p} \cdot ({\bf x} - {\bf y})}\langle \Phi_X(t,{\bf y})\ \Phi^\dagger_X(t_0,{\bf x}) \rangle \label{eq-X2pt} \\
C^{\alpha}_{B_qX}(t_0,t,T; {\bf p}) &=& \frac{1}{L^3} \sum_{{\bf x}, {\bf y}, {\bf z}} e^{i\,{\bf p}\cdot ({\bf z} - {\bf x})} \langle \Phi_X(t_0+T,{\bf x})\ J(t,{\bf z})\ \Phi^{\alpha\dagger}_{B_q}(t_0,{\bf y}) \rangle \label{eq-3pt}
\end{eqnarray}
In three-point correlator data the parent meson is created at time-slice $t_0$, the daughter meson is annihilated at $t_0+T$, and a flavor-changing current $J(t,{\bf z})$ is inserted at intermediate times $t_0 \leq t \leq t_0+T$, where $t_0$ is chosen at random to reduce auto-correlations.  This three-point correlator setup is depicted in Fig.~\ref{fig-feyndiag}.  Data are generated over the ranges of parent and daughter meson temporal separations listed in Table~\ref{tab-ens}.  Prior to fitting, all data are shifted to a common $t_0=0$.

\subsection{Fitting Correlator Data}
\begin{figure}[t]
\vspace{-0.05in}
\centering
\hspace{0.0in}  
\subfloat[][Fits to $aM_{B_s}^{(0)}$ for various source-sink combinations.]
{\scalebox{0.9}{\includegraphics[angle=0,width=0.5\textwidth]{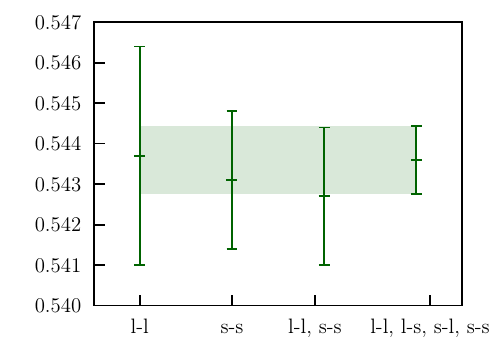}}}
\hspace{0.2in}  
\subfloat[][Fits to $b_s^{{\rm l},(0)}$ for various source-sink combinations.]
{\scalebox{0.9}{\includegraphics[angle=0,width=0.5\textwidth]{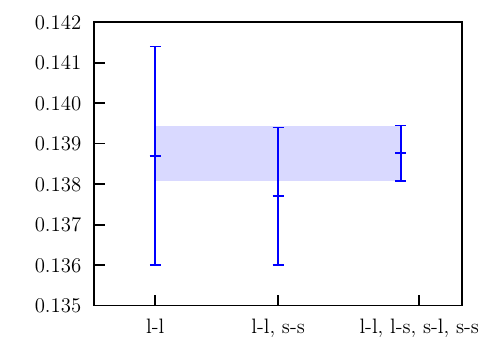}}} 
\\
\hspace{0.0in}  
\subfloat[][$aM_{B_s}^{(0)}$ vs. the number of states $N$, with $\nicefrac{t_{\rm min}}{a}=2$.]
{\scalebox{0.9}{\includegraphics[angle=0,width=0.5\textwidth]{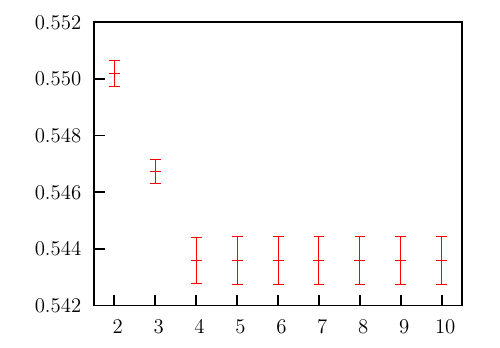}}}
\hspace{0.2in}  
\subfloat[][$aM_{B_s}^{(0)}$ versus $\nicefrac{t_{\rm min}}{a}$, with $N=10$.]
{\scalebox{0.9}{\includegraphics[angle=0,width=0.5\textwidth]{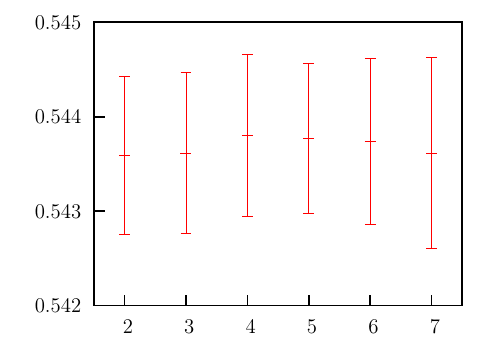}}}
\caption{From ensemble C2:  (a) and (b) show the improvement from simultaneous fits to local (``l'') and smeared (``s'') source-sink combinations; (c) and (d) display stability of fit results.}
\vspace{-0.2in}
\label{fig-parentfits}
\end{figure}
\begin{figure}[t]
\vspace{0.0in}
\begin{minipage}[b]{0.43\linewidth}
\centering
{\scalebox{1.05}{\includegraphics[angle=0,width=1.0\textwidth]{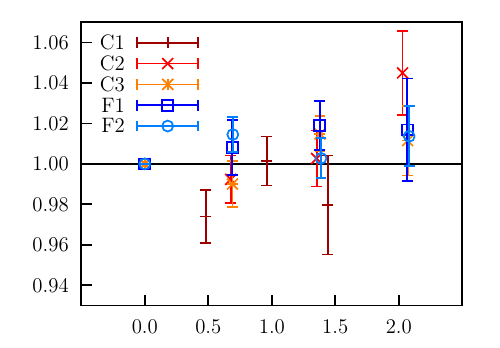}}}
\vspace{0.0in}
\caption{Fit results for $E_\pi$ and $M_\pi$ are combined with simulated pion momentum to check the dispersion relation.  A plot of $(E_\pi^2 - M_\pi^2) / {\bf p}^2$ vs. $(r_1{\bf p})^2$ is consistent with the expected result of $1+\mathcal{O}(a{\bf p})^2$.}
\label{fig-dispreln}
\end{minipage}
\hspace{0.1in}
\begin{minipage}[b]{0.55\linewidth}
\centering
{\scalebox{1.05}{\includegraphics[angle=0,width=1.0\textwidth]{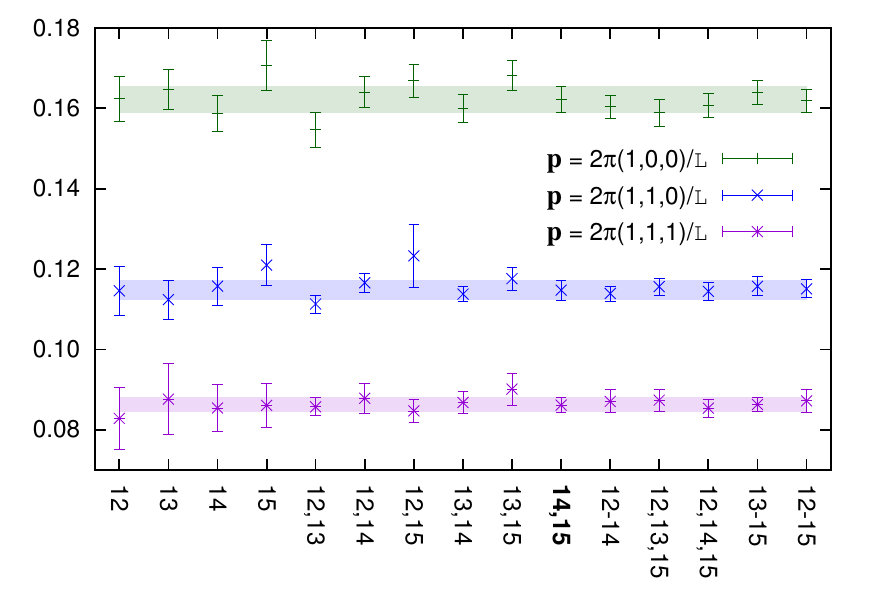}}}
\caption{For $B\to\pi\ell\nu$ on ensemble C3, fit results for $a\langle V_k \rangle_{\rm cont}$ (defined in Sec.~\protect\ref{sec-match}) are shown for combinations of $T$ used in simultaneous fits.   Colored bands correspond to the ``best-fit'' combination $T=14, 15$.}
\label{fig-3ptfits}
\end{minipage}
\end{figure}

 
Two-point correlator data for parent mesons are fit to the ansatz
\begin{equation}
C^{\alpha \beta}_{B_q}(t) = \sum^{N-1}_{n=0} b_q^{\alpha (n)} b_q^{\beta (n) \dagger} (-1)^{nt} e^{-M_{B_q}^{(n)}t}, \hspace{0.2in} {\rm where} \hspace{0.2in} b_q^{\alpha(n)} = \frac{ \langle \Phi_{B_q}^\alpha | B_q^{(n)} \rangle }{ \sqrt{2 M_{B_q}^{(n)}} }.
\end{equation}
Data are generated and analyzed for the $B$ and $B_s$ mesons and all four combinations of local and Gaussian smeared sources and sinks.  Fig.~\ref{fig-parentfits} shows the improvement observed from simultaneously fitting multiple source-sink smearing combinations and the stability of fit results with respect to changes in $N$ and $t_{\rm min}$.  Two-point correlator data for the daughter mesons are fit to
\begin{equation}
C_X(t) = \sum^{N-1}_{n=0} \big| d_X^{(n)} \big|^2 (-1)^{nt} \left( e^{-E_X^{(n)}t} + e^{-E_X^{(n)}(T-t)} \right), \hspace{0.2in} {\rm where} \hspace{0.2in} d_X^{(n)} = \frac{ \langle \Phi_X | X^{(n)} \rangle }{ \sqrt{2 E_X^{(n)}} }.
\end{equation}
We generate and analyze data for the $\pi$, $K$, and $\eta_s$ daughter mesons, each at momenta ${\bf p} \in \nicefrac{2\pi}{L}\times \{(0,0,0), (1,0,0), (1,1,0), (1,1,1)\}$.  These fit results satisfy the dispersion relation as shown in Fig.~\ref{fig-dispreln}.
At each momentum, three-point correlator data are fit to
\begin{equation}
C^\alpha_{B_qX}(t,T) = \sum_{m,n=0}^{N-1} b_q^{\alpha(m)}\ A_{B_qX}^{(m,n)}\ d_X^{(n)\dagger}\ (-1)^{mt+n(T-t)}\ e^{-M_{B_q}^{(m)}t} e^{-E_X^{(n)}(T-t)},
\end{equation}
where the three-point amplitude is related to the lattice matrix element by
\begin{equation}
A_{B_qX}^{(n,m)} = \frac{\langle X | J | B_q \rangle}{2\sqrt{M_{B_q}^{(n)} E_X^{(m)}} }.
\label{eq-lattJ}
\end{equation}
We perform a simultaneous, Bayesian fit to the four local and smeared combinations of the parent two-point, the daughter two-point, and three-point correlator data sets for multiple values of $T$.  The improvement from simultaneously fitting data for multiple $T$ is shown in Fig.~\ref{fig-3ptfits}.

\subsection{Matching and Preliminary Results}
\label{sec-match}
The lattice vector current ($J = \mathcal{V}_\mu$) is matched to the continuum at one-loop using massless HISQ lattice perturbation theory~\cite{HPQCD:2006, HPQCD:renorm}
\begin{equation}
\langle V_\mu\rangle_{\rm cont} = (1+\alpha_s \rho_\mu^{(0)})\langle \mathcal{V}_\mu^{(0)}\rangle + \langle \mathcal{V}_\mu^{(1), {\rm sub}}\rangle
\end{equation}
where $\langle \mathcal{V}_\mu^{(1), {\rm sub}}\rangle \equiv \langle \mathcal{V}_\mu^{(1)}\rangle - \alpha_s \zeta_{10, \mu} \langle \mathcal{V}_\mu^{(0)}\rangle$.  Currents contributing through $\mathcal{O}(\alpha_s, \nicefrac{\Lambda_{\rm QCD}}{M}, \nicefrac{\alpha_s}{aM})$ are
\begin{equation}
\mathcal{V}_\mu^{(0)} = b\ \gamma_\mu\ \bar{q} \hspace{0.3in} {\rm and} \hspace{0.3in} \mathcal{V}_\mu^{(1)} = -\frac{1}{2M}\ b\ \gamma_\mu\, {\boldsymbol \gamma} \cdot {\boldsymbol \nabla}\ \bar{q}
\end{equation}
where $q$ is the daughter quark in Fig.~\ref{fig-feyndiag}.  For the lattice tensor current ($J = \mathcal{T}_{\mu\nu}$),
\begin{equation}
\langle T_{k0} \rangle_{\rm cont} = ( 1 + \alpha_s \rho_T ) \langle \mathcal{T}_{k0}^{(0)} \rangle + \langle \mathcal{T}_{k0}^{(1), {\rm sub}} \rangle
\end{equation}
where $\langle \mathcal{T}_{k0}^{(1), {\rm sub}} \rangle = \langle \mathcal{T}_{k0}^{(1)} \rangle - \alpha_s \zeta_{10}^T \langle \mathcal{T}_{k0}^{(0)} \rangle$.  Heavy-quark symmetry of the NRQCD $b$-quark allows the tensor current renormalization to be recast in terms of vector current quantities:  $\mathcal{T}_{k0}^{(0)} =  \mathcal{V}_k^{(0)}$, $\mathcal{T}_{k0}^{(1)} = -  \mathcal{V}_k^{(1)}$, and $\zeta_{10}^T = - \zeta_{10,k}$.

\begin{figure}[t!]
\vspace{0.1in}
\centering
\hspace{0.0in}  
\subfloat[][$f_{+,0}(q^2)$ for $B \to \pi \ell \nu$.]
{\scalebox{.971}{\includegraphics[angle=0,width=0.5\textwidth]{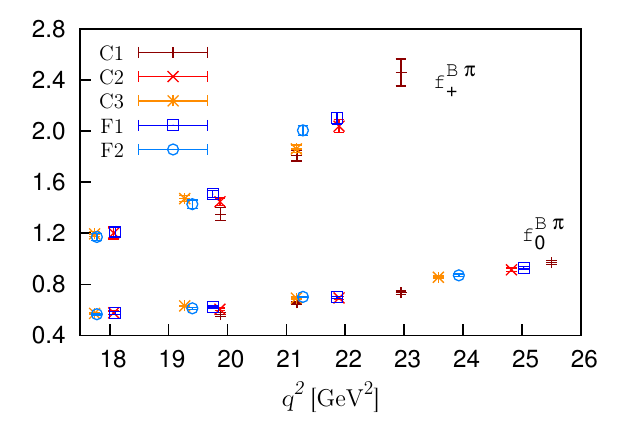}}}
\hspace{0.1in}  
\subfloat[][$f_{+,0}(q^2)$ for $B_s \to K \ell \nu$.]
{\scalebox{.971}{\includegraphics[angle=0,width=0.5\textwidth]{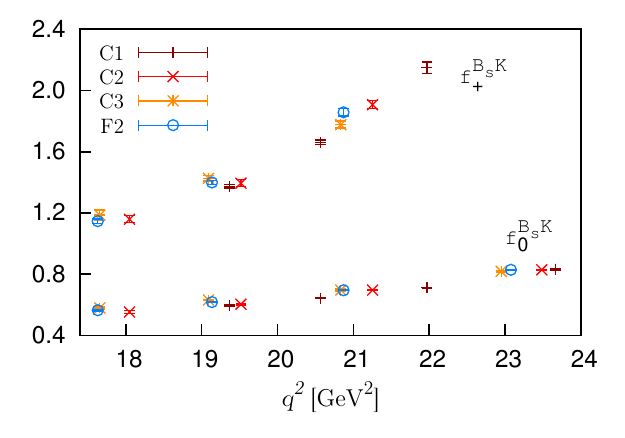}}} 
\hspace{0.0in}  
\subfloat[][$f_{+,0}(q^2)$ and $if_T(q^2)$ for $B \to K \ell^+ \ell^-$.]
{\scalebox{.971}{\includegraphics[angle=0,width=0.5\textwidth]{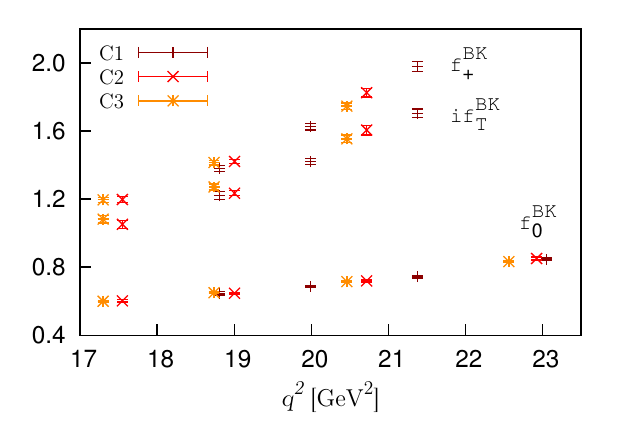}}}
\hspace{0.1in}  
\subfloat[][$f_{+,0}(q^2)$ for $B_s \to \eta_s$.]
{\scalebox{.971}{\includegraphics[angle=0,width=0.5\textwidth]{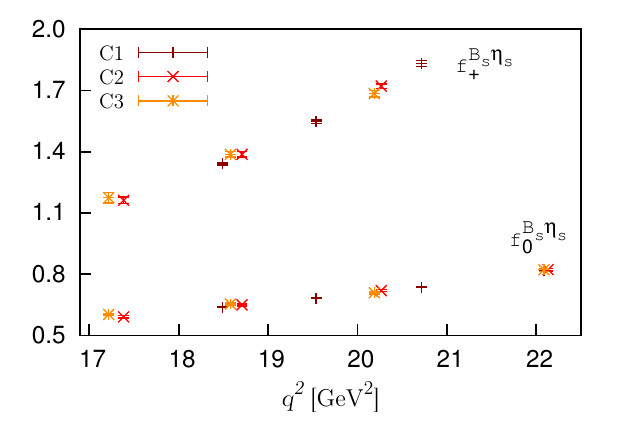}}}
\caption{Preliminary form factor results.}
\vspace{0.0in}
\label{fig-results}
\end{figure} 
For the ensembles analyzed, preliminary results for form factors are shown in Fig.~\ref{fig-results}.  The form factors $f_{+,0}(q^2)$ are calculated for all decay channels and $i f_T(q^2)$ is calculated for $B\to K\ell^+\ell^-$.

\section{Next Steps}
Once data generation and correlator fitting is complete, we will extract the physical values of the form factors.  The kinematic dependence of the form factors over the full range of physical $q^2$ can be written in a model-independent way via the $z$-expansion~\cite{zexp}.  We plan to incorporate the chiral and continuum extrapolations in the $z$-expansion as in~\cite{HPQCD:2010, HPQCD:2011}.  The resultant \emph{modified $z$-expansion} permits the use of data over the full range of $q^2$, including data at daughter momenta for which chiral perturbation theory is expected to break down.  We plan to cross-check these results against those obtained by separately performing the chiral and continuum extrapolation and then the $z$-expansion.

\section*{Acknowledgements}
Funding for this research was provided by the NSF and the the DOE.  Numerical simulations were carried out on facilities of the USQCD Collaboration funded by the Office of Science of the DOE and at the Ohio Supercomputer Center.


\end{document}